\documentclass[twocolumn,preprintnumbers,amsmath,amssymb,prb]{revtex4}

\usepackage{graphicx}
\usepackage{dcolumn}
\usepackage{bm}

\begin{document}

\title{Electronic structure and magnetism of transition metal doped Zn$_{12}$O$_{12}$ clusters: Role of defects}

\author{Nirmal Ganguli}
\affiliation{Department of Physics, Indian Institute of Technology Bombay, Mumbai 400076, India}
\affiliation{Department of Solid State Physics, Indian Association for the Cultivation of Science, Jadavpur, Kolkata 700032, India}
\author{Indra Dasgupta}
\altaffiliation[Also at ]{Center for Advanced Materials, Indian Association for the Cultivation of Science, Jadavpur, Kolkata 700032, India}
\email[Email: ]{sspid@iacs.res.in}
\affiliation{Department of Physics, Indian Institute of Technology Bombay, Mumbai 400076, India}
\affiliation{Department of Solid State Physics, Indian Association for the Cultivation of Science, Jadavpur, Kolkata 700032, India}
\author{Biplab Sanyal}
\affiliation{Department of Physics and Astronomy, Uppsala University, Box 516, SE-75120 Uppsala, Sweden}

\date{\today}

\begin{abstract}
We present a comprehensive study of the energetics and magnetic properties of ZnO clusters doped with 3$d$ transition metals (TM) using {\em ab initio} density functional calculations in the framework of generalized gradient approximation + Hubbard~U (GGA+U) method. Our results within GGA+U for all 3$d$ dopants except Ti indicate that antiferromagnetic interaction dominates in a neutral, defect-free cluster. Formation energies are calculated to identify the stable defects in the ZnO cluster. We have analyzed in details the role of these defects to stabilize ferromagnetism when the cluster is doped with Mn, Fe, and Co. Our calculations reveal that in the presence of charged defects the transition  metal atoms residing at the surface of the cluster may have an unusual oxidation state, that plays an important role to render the cluster ferromagnetic. Defect induced magnetism in ZnO clusters without any TM dopants is also analyzed. These results on ZnO clusters may have significant contributions in the nanoengineering of defects to achieve desired ferromagnetic properties for spintronic applications.
\end{abstract}
\maketitle

\section{\label{sec:intro}INTRODUCTION}
Functional materials in reduced dimension are of immense interest from the point of view of both fundamental science and technological applications as novel properties emerge due to the complex quantum mechanical interactions at this size regime. One of the major applications is in high density storage in miniaturized devices. Quantum confinement plays a key role in tailoring properties of these small systems, e.g., atomic clusters containing tens of atoms. 
In this context, magnetic clusters and nanosized systems, either in a pure atomic assembly or as magnetically doped semiconducting systems, have attracted a lot of attention. An exciting material is transition metal (TM) doped ZnO nanocrystals where ferromagnetism is reported even at high temperatures. \cite{gamelin2003, gamelin2004, kittilstved, martinez, karmakar, kataoka} This system has received special attention as naturally abundant environment friendly ZnO with its wide band gap coupled with magnetism may provide a unique opportunity to achieve electronic, magneto-electronic, and opto-electronic multifunctionality in a single system.

First principles electronic structure calculations on TM doped ZnO clusters have been performed to provide useful insights regarding the electronic structure and magnetic interactions. Reber~{\em et~al.} have studied a small cluster of ZnO (${\rm Zn_{12}O_{12}}$) in the framework of generalized gradient approximation (GGA) and reported that doping of Co leads to ferromagnetic coupling due to direct exchange interaction in this system \cite{reber}. It was also reported that doping of Mn into the ${\rm Zn_{12}O_{12}}$ cluster yields antiferromagnetic (AFM) ground state for small Mn-Mn distance while ferromagnetic and AFM states are degenerate for large Mn-Mn distance \cite{liu}. Very recently the structural and magnetic properties of several 3$d$ TM doped Zn$_{12}$O$_{12}$ clusters were investigated in the framework of GGA.\cite{chen} The authors find that most of the TM dopants stabilize either in paramagnetic or in antiferromagnetic state, except Co and Ni, where the ferromagnetic and antiferromagnetic states are found to be energetically very close.

The experimental works on TM doped ZnO clusters are primarily focused on the study of ferromagnetism (FM) in TM doped ZnO nanoparticles.\cite{gamelin2003, gamelin2004, kittilstved, martinez, karmakar, kataoka} Recently Karmakar {\em et~al.} have found ferromagnetism in Fe doped ZnO nano-crystals.\cite{karmakar} However for these systems the local probes like electron paramagnetic resonance (EPR) and electron spin resonance (ESR) reveal an unusual ionic state of the dopant Fe atoms. They find the presence of both: Fe$^{2+}$ and Fe$^{3+}$ ionic states in the system and argue that these are crucial for ferromagnetism. The presence of Zn vacancies are speculated to play an important role in stabilizing Fe$^{3+}$ ions in the system. X-ray Magnetic Circular Dichroism (XMCD) studies on the same samples later confirmed that Fe$^{3+}$ ions are located at the surface of the nanoparticles and are crucial for their magnetic properties.\cite{kataoka} It has been reported that the type of carriers also play an important role in ferromagnetism in TM doped ZnO quantum dots: Mn$^{2+}$ doped ZnO quantum dots become ferromagnetic in presence of $p$-type carriers, while $n$-type carriers induce ferromagnetism in Co$^{2+}$ doped ZnO quantum dots.\cite{kittilstved} Interestingly an experimental report indicates that ferromagnetism in ZnO and in many other oxide nanoparticles may be realized even in the absence of TM doping presumably promoted by defects.\cite{sundaresan} The possibility of inducing room temperature ferromagnetic-like behavior in ZnO nanoparticles without doping magnetic impurities but coating with N and S containing ligands such as dodecylamine and dodecanethiol has also been realized.\cite{garcia}

The above discussion clearly suggests that magnetism in TM doped ZnO clusters and nanosystems is rather subtle, where surface effects and defects play a crucial role to stabilize FM. Therefore, an understanding of the defect induced magnetism is crucial to design these materials for desirable property.
In a recent work using GGA+U on Fe doped ZnO clusters we have shown that defects under suitable conditions can induce ferromagnetic interactions between the dopant Fe atoms, whereas antiferromagnetic coupling dominates in a neutral defect-free cluster.\cite{ganguli} In addition, there are several calculations that suggest that ZnO nanocluster may be ferromagnetic without TM dopants. Schoenhalz~{\em et~al.} have proposed that the surface states promoted by extended defects ({\em e.g.} grain boundaries) may play an important role in mediating ferromagnetic interaction in such materials.\cite{schoenhalz} A recent first principles study of ZnO nanostructures coated with ligands containing N and S also reported ferromagnetism due to the redistribution of charge promoted by ligand capping.\cite{wang}

In this paper we shall present a systematic study and analysis of the electronic structure and magnetism of transition metal doped Zn$_{12}$O$_{12}$ cluster in the framework of GGA+U. We have carried out our studies both in the presence as well as in the absence of defects in order to clarify the nature and origin of magnetic interactions. Our study along the 3$d$ transition metal series will be useful to extract trends. Our calculations demonstrate that doping TM except Ti in pristine ZnO does not render the cluster ferromagnetic, however defects under suitable condition can induce ferromagnetic interaction between the dopant TM ions.

The remaining part of the paper is organized as follows: In section~\ref{sec:methodology} we have described the technical details and the method of our calculations. Details about obtaining the structure of the cluster is given in section~\ref{sec:structure}. We have discussed the results for doping of TM into the cluster in section~\ref{sec:gga+u}. Effect of defects namely Zn and O vacancies with different charge states in the presence and in the absence of TM dopants are discussed in sections~\ref{sec:vacancy} and \ref{sec:vacancyOnly} respectively. Finally, we have summarized our work in section~\ref{sec:conclusion}.

\section{\label{sec:methodology}Methodology}
The total energy and electronic structure calculations presented in this paper are carried out in the framework of {\em ab initio} density functional theory (DFT). We have used plane wave basis along with the projector augmented wave (PAW) method \cite{blochl} as implemented in the VASP\cite{vasp2, vasp} code. The exchange-correlation term in DFT was treated within the generalized gradient approximation (GGA) due to Perdew-Burke-Ernzerhof (PBE).\cite{pbe} For the description of dopant TM atoms we have employed GGA+U with the typical value of on-site $d-d$ Coulomb interaction $U=4.0$~eV and the onsite exchange interaction $J=1.0$~eV. PAW potentials with 12 valence electrons ($3d^{10}4s^2$) for Zn, 6 ($2p^42s^2$) for O, and $2+n$ ($3d^n4s^2$); $n=2$ to 9 for TMs (from Ti to Cu) have been used. The plane wave energy cutoff was taken to be 500~eV. The cluster was simulated in a large cell (a cube of volume (20~\AA)$^3$) to ensure sufficient separation between the periodic images. Only one $k$-point ({\em i.e.} $\Gamma$ point) was used for these calculations. We have relaxed the atomic positions to minimize the Hellman-Feynman forces on each atom with the tolerance value of 0.01~eV/\AA.

The formation energy (FE) of defects is the energy for the reaction needed to create the defect from the ideal material. It is calculated using the following expression \cite{ganguli}:
\begin{eqnarray}
{\rm FE}&=&E{\rm [Zn_mO_n}(\alpha, q)] - E{\rm [Zn_pO_p(pure)]} \nonumber \\
&+& \sum_{\alpha} n_{\alpha}\mu_{\alpha} + q(E_v + \epsilon_F),
\label{eq:fe}
\end{eqnarray}
where $\alpha$ is the defect atom added to or removed from the pristine ZnO cluster, $n_{\alpha}$ is the number of each defect atom: $n_{\alpha}=-1~(+1)$ for adding (removing) {\em one} atom, $E{\rm [Zn_mO_n}(\alpha, q)]$ is the total energy with defect $\alpha$ and charge $q$, while $E{\rm [Zn_pO_p(pure)]}$ is the total energy of pure ZnO cluster. $\epsilon_F$ is the Fermi level measured with respect to the energy ($E_v$) of the highest occupied molecular orbital (HOMO) in pristine ZnO cluster.  $\mu_{\alpha}$ is the chemical potential of atom $\alpha$ in some suitable reservoir.

The chemical potential of a particular atom is the energy supplied (released) to add (remove) {\em one} atom of that type to (from) the system. It depends on the conditions under which the material has been prepared. However, there are limits on the values of chemical potentials of Zn and O for the formation of ZnO cluster. In equilibrium the chemical potentials of all Zn and O atoms should add up to the total energy of the cluster, {\em i.e.} the following equation must be obeyed:
\begin{eqnarray}
p(\mu_{\rm Zn} + \mu_{\rm O}) = E[{\rm Zn_pO_p(pure)}],
\label{eq:constraint1}
\end{eqnarray}
where $p$ is the number of ZnO units in the cluster. However, the chemical potential of an individual atom can not be more than its standard value in elemental structure because otherwise the standard elemental structure will be more stable and in that case the cluster will never be formed. This imposes the following constraints on the chemical potentials:
\begin{eqnarray}
\mu_{\rm Zn} \leq \mu^s_{\rm Zn} ~{\rm and}~ \mu_{\rm O} \leq \mu^s_{\rm O},
\label{eq:constraint2}
\end{eqnarray}
where $\mu^s_{\rm Zn}$ and $\mu^s_{\rm O}$ are chemical potentials in standard elemental structures of Zn and O respectively.
Both the above constraints can be simultaneously satisfied if $\mu^s_{\rm Zn}+\mu^s_{\rm O} \geq E[{\rm Zn_pO_p(pure)}]/p$, and Eq.~(\ref{eq:constraint1}) can be rewritten in the following way:
\begin{eqnarray}
p(\mu^s_{\rm Zn} + \mu^x_{\rm Zn}+ \mu^s_{\rm O}+\mu^x_{\rm O}) = E[{\rm Zn_pO_p(pure)}],
\label{eq:constraint}
\end{eqnarray}
where $\mu^x_{\rm Zn}$ is the excess chemical potential of Zn, defined as: $\mu^x_{Zn} = \mu_{Zn} - \mu^s_{Zn}$. Excess chemical potential of O ($\mu^x_{\rm O}$) is also defined in the similar way. We gather from Eq.~(\ref{eq:constraint2}) that the conditions: $\mu^x_{\rm Zn} \leq 0$ and $\mu^x_{\rm O} \leq 0$ must be satisfied and
Eq.~(\ref{eq:constraint}) indicates that there is only one free parameter between $\mu^x_{\rm Zn}$ and $\mu^x_{\rm O}$ and there are lower limits on the values of these excess chemical potentials. Here we have taken two extreme limits of chemical potentials to calculate the formation energies: (i) Zn rich limit ($\mu^x_{\rm Zn} = 0$, $\mu^x_{\rm O}$ is minimum) and (ii) O rich limit ($\mu^x_{\rm O} = 0$, $\mu^x_{\rm Zn}$ is minimum). The chemical potential in standard elemental structure of atom $\alpha~(\mu^s_\alpha)$ was taken to be the energy per atom of the element in the most stable elemental structure (bulk solid or molecular form).
We have simulated different defect charge states by adding or removing electrons from the system. The additional charge was compensated by a uniform jellium background suitable for the system to ensure charge neutrality of the system. In view of the large supercell chosen in our calculation, we have not included any finite size correction for the charged systems.\cite{yu}

\section{\label{sec:structure}Structure of the cluster}

We have considered Zn$_{12}$O$_{12}$ cluster for our simulations. The cluster of this particular size has been reported to be the magic size cluster with the most stable configuration.\cite{reber, yadav} In this context, it is interesting to note that the experimental absorption energies of small ZnO nanoparticles are found to be very close to the optical excitation energy of Zn$_{12}$O$_{12}$ cluster calculated in the framework of time dependent density functional theory (TDDFT).\cite{matxain} This observation leads the authors of Ref.~\onlinecite{matxain} to conjecture that the surface structure of small ZnO nanoparticles is very similar to that of Zn$_{12}$O$_{12}$ cluster. In view of the above discussion, our study of TM doped Zn$_{12}$O$_{12}$ cluster may help understanding the role of surface in the magnetic properties of TM doped ZnO nanoparticles.

We have considered 12 ZnO units in several different initial configurations and relaxed the atomic positions to obtain the ground state structure by minimizing the forces on the atoms. The lowest energy structure containing all the atoms arranged on the surface of a hollow sphere is displayed in Fig.~1.
\begin{figure}
\centering
\includegraphics[scale = 0.45]{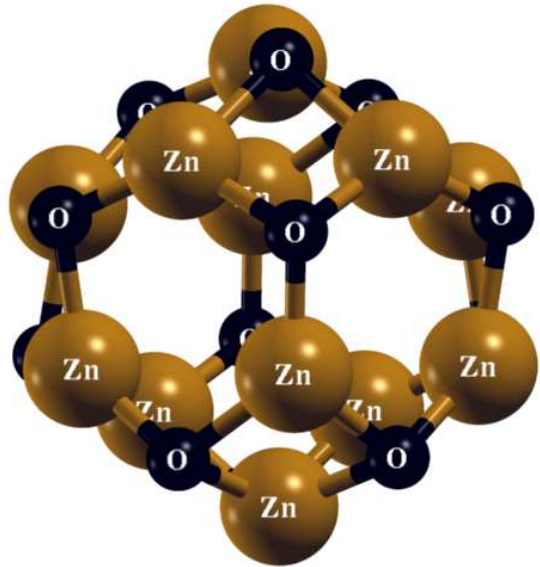}
\caption{Optimized structure of Zn$_{12}$O$_{12}$ cluster.}
\end{figure}
Each atom of the cluster is coordinated with three neighbors and the atoms are arranged periodically with 6 four-atom rings and 8 six-atom rings. This structure is in good agreement with the structure reported earlier for the same size of the cluster \cite{reber, yadav, liu, matxain}.
The diameter of the cluster is $\sim$6.35~\AA. 
Although the size of the cluster is much smaller than the nanoparticles synthesized experimentally, however since all the atoms are located on the surface of a sphere in this cluster, it provides a unique opportunity to study in details the surface contribution to magnetism that will be relevant for a ZnO  nanoparticle.
The gap between HOMO and the lowest unoccupied molecular orbital (LUMO) is calculated to be $\sim$2.34~eV.

\section{\label{sec:result}Results and Discussions}
As a next step, we consider doping of the cluster with TM ions. In this context, recent theoretical studies on TM doped bulk ZnO indicate that standard exchange-correlation (XC) functionals such as local spin density approximation (LSDA)/GGA are in general not adequate to describe these systems. One such report \cite{chanier} points out that the LSDA predictions might be misleading as the localized character of TM impurities are not taken into account. On the contrary, incorporation of Hubbard U gives more appropriate description of the system \cite{spaldin}. Very recently, Iu\c{s}an et al.\cite{iusan09} have shown that the  magnetic exchange parameters calculated with LSDA+U for Co doped ZnO are in good agreement with the magnetic measurements.\cite{sati} Considering the importance of Coulomb correlation, we have employed GGA+U method for the description of TM-$d$ states in the TM doped ZnO cluster. We have calculated the stability of the cluster on doping, the corresponding magnetic moments, and magnetic interactions between two TM ions. The positions of the atoms in the cluster were fully relaxed upon doping.

\subsection{\label{sec:gga+u}Doping of TM: GGA+U results}
We have only simulated substitutional doping in view of a recent report on Mn doped ZnO cluster where it is found that substitutional doping of Mn at the Zn site is energetically more favorable compared to endohedral or exohedral doping \cite{liu}.
From our calculations we find that doping of TM at the Zn site is energetically more favorable than doping it at the O site. Therefore we have considered only substitutional doping of TM at Zn site for our work.

We have calculated the formation energies of doping one TM ion into the system using Eq.~(\ref{eq:fe}) both at Zn rich and O rich limits. Magnetic moments corresponding to doping of different TMs are also calculated. The results of our calculations are presented in Fig.~2.
\begin{figure}
\centering
\includegraphics[scale = 0.40]{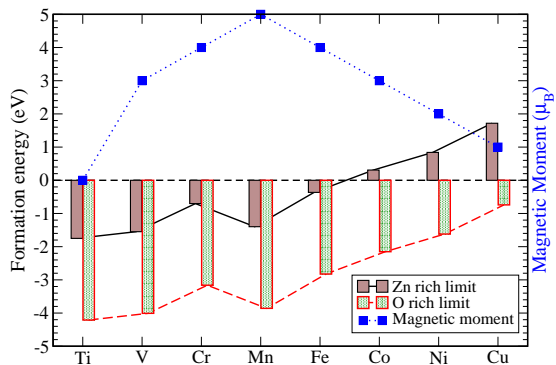}
\caption{Formation energies (in Zn rich and O rich limit) and magnetic moments are shown for Zn$_{12}$O$_{12}$ cluster doped with different TMs.}
\end{figure}
We find that the Zn rich limit is hardly favorable for TM doping. However, as expected, at O rich limit formation energies corresponding to doping of all TM ions are negative, which implies that O rich limit is more favorable for doping of TM into the cluster. In practice, doping of Ti in ZnO at the O rich limit may not be achieved easily as there is a possibility of formation of other oxides of Ti in this limit.

The magnetic moments of the clusters show interesting features depending upon the dopant TM ions. As we move from Ti to Cu, the magnetic moment increases first, becomes maximum $(5~\mu_B)$ for Mn and then decreases monotonically upto Cu. The magnetic moments are mainly localized on the TM ions. The TM ions are expected to be in $2+$ oxidation state since they substitute Zn$^{2+}$ ion. The values of magnetic moments can be explained by Hund's rule {\em i.e.} assuming high spin arrangement of $3d$ electrons for TM$^{2+}$ ions, with the exception of Ti. The magnetic moment is completely quenched upon Ti doping.
It is interesting to note that, as opposed to other TM dopants, the Ti ion and the three neighboring oxygens surrounding Ti are confined in a plane upon relaxation. The Ti-O bond lengths are found to be quite small (1.86, 1.86, and 1.89~\AA) compared to other TM-O bond lengths ({\em e.g.} Mn-O bond lengths are 1.95, 2.04, and 2.04~\AA). Therefore, the crystal field experienced by the Ti ion is stronger compared to that for the other TM dopants. This crystal field is in fact stronger than the exchange splitting for Ti, thus stabilizing the low spin state for the singly doped Ti ion. 

The spin polarized densities of states corresponding to doping of {\em one} TM ion into the ZnO cluster are shown in Fig.~3.
\begin{figure}
\vskip 0.5cm
\centering
\includegraphics[scale=0.45]{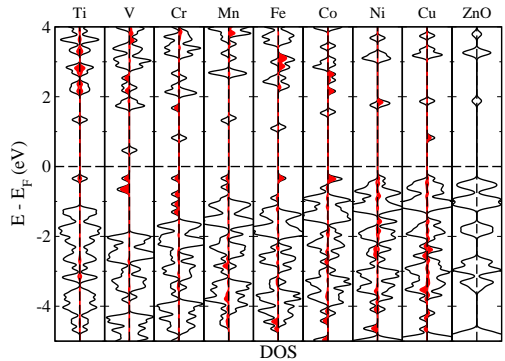}
\caption{Total density of states (DOS) and partial density of states (PDOS) corresponding to TM-$d$ states for Zn$_{12}$O$_{12}$ clusters doped with one TM. Partial Density of States of TM-d states are shaded, whereas solid lines indicate total density of states for ZnO cluster doped with {\em one} TM. For each vertical panel, sub-panel of left hand side indicates majority spin states and sub-panel of right hand side indicates minority spin states. DOS of pristine ZnO cluster has also been shown at the rightmost panel for comparison. Gaussian smearing scheme with 0.1~eV of smearing width was used for the DOS calculations.}
\end{figure}
The DOS corresponding to the pristine ZnO cluster has also been shown at the rightmost panel for comparison. The figure shows that the TM-d states are lying deep into the gap region. We gather from the figure that as expected for high spin TM$^{2+}$ ions, the minority $d$-states are completely unoccupied for V, Cr, and Mn, while for Fe, Co, Ni, and Cu the minority states are progressively filled, accounting for the reduction in magnetic moment. Further, the overlap of the TM-$d$ states with the host states increase as we move from Ti to Cu.

After exploring the possibility of doping of TM into the cluster, we have calculated the magnetic exchange interaction between TM ions in this system. To study the exchange interaction between TM ions it is essential to substitute at least {\em two} TM ions into the cluster. We have identified 7 different possible distances between Zn atoms, where a pair of TMs can be doped in pristine ZnO. Hence we have got 7 different configurations with increasing distances.
Doping of TM ions in all possible configurations are assumed and the atomic positions are relaxed separately for parallel and antiparallel orientations of magnetic moments on TM ions. The energy differences between antiparallel (AFM) and parallel (ferromagnetic) states for different configurations have been plotted as a function of transition metal and displayed in Fig.~4.
\begin{figure}
\vskip 0.5cm
\centering
\includegraphics[scale = 0.45]{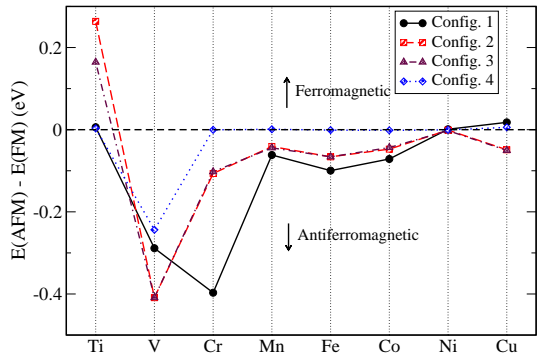}
\caption{Exchange interaction as a function of dopant TM for first four configurations.}
\end{figure}
Since for higher configuration numbers (larger distances between TM ions) the exchange interactions effectively die out, here we have only shown the interactions upto fourth configuration. The fast decay of the interaction strength with distance may be attributed to the following reason: It has been observed that TM-$d$ states are mainly located deep inside the HOMO-LUMO gap region (see Fig.~3). Since these states are not delocalized over the cluster, they can only interact with a state located very close to it in space, accounting for the sharp decay.

We find that except Ti, the dominant exchange interaction for all the other transition elements (V-Cu) doped into ZnO cluster to be antiferromagnetic. This is in sharp contrast to the GGA results discussed in the literature,\cite{reber, liu, chen} where one finds the tendency of ferromagnetism in Co, Cu, and Ni doped ZnO cluster and the system is non-magnetic upon Ti doping. Therefore our calculation emphasizes the importance of inclusion of Hubbard U in the study of magnetism in TM doped ZnO cluster.

Fig.~4 reveals that the exchange interaction for a pair of Ti atoms is ferromagnetic. It has a magnetic moment of 2~$\mu_B$. We have seen that Ti has a large crystal field splitting and weak exchange splitting. However, a pair of Ti atoms may interact forming bonding and antibonding states and this splitting is rather strong for Ti in comparison to other TM atoms doped into the system. In contrast to other TM's, a pair of Ti atoms are found to come closer to each other upon relaxation. 
For example, the separation between Ti atoms for the first three configurations are 2.44, 2.66, and 2.75~\AA    respectively, while for Mn it is 2.91, 3.37, and 3.38~\AA.
This provides strong bonding-antibonding interaction between similar spin states promoting an arrangement of majority and minority states as shown in Fig.~5 (partial DOS corresponding to Ti-$d$ states), accounting for the calculated magnetic moment.
\begin{figure}
\includegraphics[scale = 0.45]{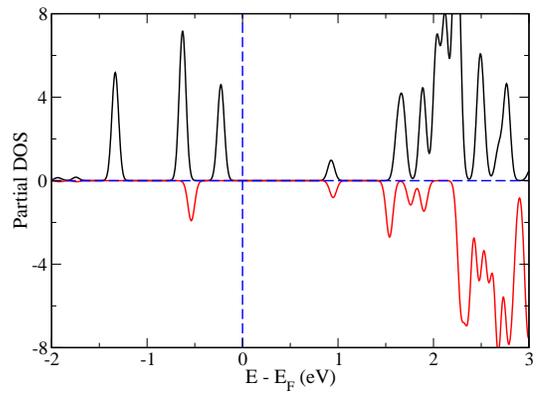}
\caption{Partial DOS corresponding to Ti-$d$ states near the Fermi level for Ti$_2$Zn$_{10}$O$_{12}$ cluster.}
\end{figure}
\subsection{\label{sec:vacancy}Effect of vacancy}
From the results of the preceding section we gather that ferromagnetism is very unlikely to be stable in transition metal doped pristine ZnO clusters with the possible exception of Ti. However, there are several reports that claim room temperature ferromagnetism in Mn, Fe and Co doped ZnO, which suggests that defects may play an important role in stabilizing ferromagnetism in such systems. Different types of native defects ({\em e.g.} interstitials, vacancies, and antisites) have been identified for ZnO in bulk form. It was reported from first-principles calculations that Zn and O vacancies are the most relevant defects in ZnO \cite{kohan}.

In view of the above we have studied the TM doped Zn$_{12}$O$_{12}$ clusters in the presence of defects. Here we have only considered Mn, Fe, and Co as dopants, since majority of the experimental results on ZnO nanosystems are reported with these dopants.

To find out the stable defect states in TM doped ZnO cluster, we study the formation energies for a single transition metal doping with and without Zn and O vacancy in different charge states. Similar to the previous cases, we have calculated formation energies in Zn and O rich limit.
We have studied two different configurations depending on the distance between TM ion and vacancy, namely near (TM ion and vacancy are at the closest possible positions) and far (distance between TM ion and vacancy is maximum). We considered both Zn rich and O rich conditions. The Fermi level has been varied from the HOMO level of the pristine ZnO cluster to the value of the experimentally reported band gap of bulk ZnO.

The formation energies for doping of Mn, Fe, and Co into the ZnO cluster with and without vacancies at different charge states are shown in Fig.~6. Since the formation energies in the Zn rich limit is never found to be lower than that in the O rich limit, here we have only displayed our results in the latter limit. Fig.~6 shows that TM doped ZnO in the charge state +1 has the lowest formation energy at the $p$-type regime, whereas Zn vacancy in the charge state $-1$ is the most stable state at the $n$-type regime. 
Here we note that the stable defects may
lead to 3+ oxidation state of the TM atom on the surface of the cluster. This is in agreement with a recent experimental report on Fe doped ZnO nanocrystals \cite{kataoka}. The unusual ionic state may be attributed to different coordination of TM atoms on the surface of the cluster compared to the bulk terminated ones due to the loss of ligands.

Having established the most stable defects in the TM doped cluster, we have investigated whether the most stable configuration may support the tendency of FM. The energy differences between the AFM and ferromagnetic configurations with the inclusion of two TM ions in the absence and presence of Zn and O vacancies in various charge states are shown in Fig.~7. Here the TM ions are always assumed to be closest to each other. Calculations are carried out when the defects (Zn or O vacancy) are near as well as far from the pair of transition metals (TM) doped into the cluster. From Fig.~7 we gather that in the absence of defects Mn, Fe, Co doped ZnO cluster in the charge state $+1$ exhibit tendency for ferromagnetism.  Zn vacancies in the $-1$ charge state also induces a tendency of FM into the cluster for all three dopant transition elements (Mn, Fe, and Co). Although Zn vacancy in +1 charge state shows tendency of ferromagnetism upon doping of Mn, this result is not interesting as Zn vacancy in +1 charge state is not stable. We note that in either case, TM doped ZnO cluster in the charge state +1 without any defects or in the charge state -1 in the presence of Zn vacancy leads to 3+ oxidation state for the dopant TM and are important to promote ferromagnetism.

\begin{figure}
\includegraphics[scale=0.45]{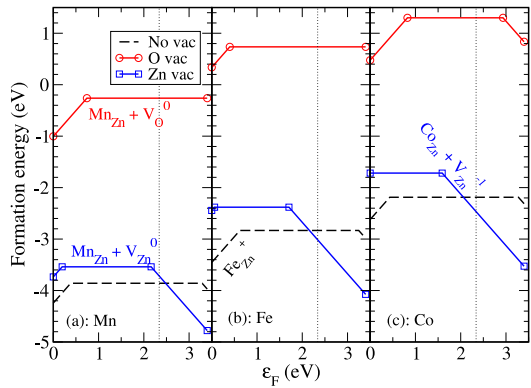}
\caption{\label{fig:1tmVacForm}(Color online) Calculated formation energies for single TM atom doped cluster at oxygen rich limit with and without vacancies in different charge states. Only the stable part of a particular charged state is shown. When vacancy is present, TM atom and vacancy are at the closest possible positions.
The Fermi energy $\epsilon_F$ has been varied upto experimental band gap of bulk ZnO and dotted vertical line indicates the calculated band gap for pure cluster.
}
\end{figure}

\begin{figure}
\includegraphics[scale=0.43]{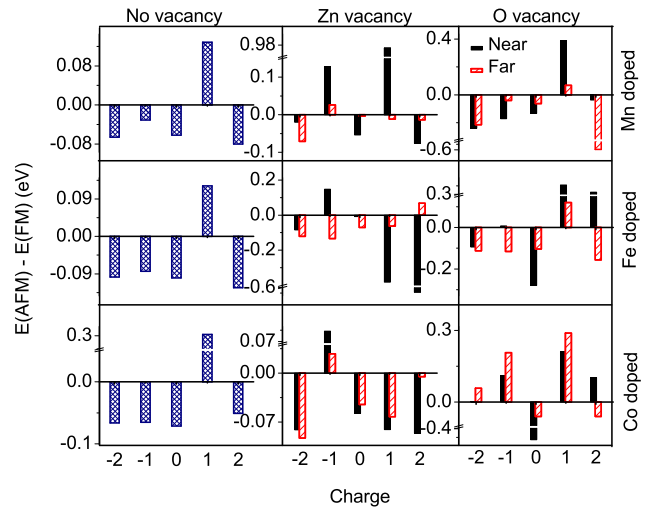}
\caption{Calculated energy differences between AFM and FM orientations for different charge states. Positive (negative) value indicates ferromagnetic (antiferromagnetic) ground state.}
\end{figure}

In order to understand the possible mechanism that induces ferromagnetic order, we consider a representative case namely a cluster doped with 2 Mn ions along with one Zn vacancy in the $-$1 charge state. The presence of a Zn vacancy in the Zn$_{12}$O$_{12}$ cluster doped with a pair of Mn atoms in the charge state -1 relaxes significantly, as shown in Fig.~8(a). In the present case the oxygen atoms surrounding the Zn vacancy are found to undergo large ($\sim$28\%) outward relaxation. Similar trend is seen for the other stable charge states. Next we shall address the magnetism of the cluster. In the presence of a Zn vacancy in the charge state $-1$ there is one hole in the system.
\begin{figure}
\includegraphics[scale = 0.45]{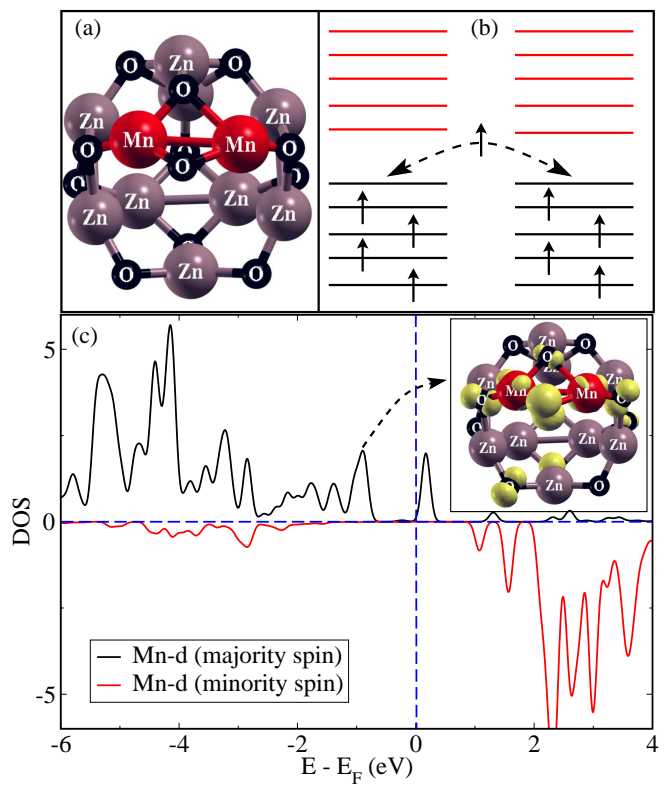}
\caption{ZnO cluster doped with 2 Mn ions along with one Zn vacancy and $-$1 charged state, (a): relaxed structure of ZnO cluster doped with two Mn atoms and one zinc vacancy in the charge state -1, (b): level diagram showing expected interaction between $d$ states of two Mn atoms (energy levels in black correspond to majority spin states and energy levels in red correspond to minority spin states), and (c): partial DOS for Mn-d states (inset shows the charge density corresponding to highest occupied Mn-d (majority) state).}
\end{figure}
When Mn ions are substitutionally doped at Zn sites, its oxidation state should be Mn$^{2+}$. Due to deficiency of one electron in the system, one Mn ion may be in 3+ oxidation state resulting in mixed valency with a combination of Mn$^{2+}$ and Mn$^{3+}$ states present in the system. In such a situation if the spins are ferromagnetically aligned as shown in Fig.~8(b) then there is a possibility of lowering energy by spin conserved hopping as illustrated in Fig.~8(b).
To verify the above mentioned mechanism, we have plotted partial DOS for Mn-d states in Fig.~8(c) and the charge density corresponding to the highest occupied Mn-d (majority) state in Fig.~8(c)(inset). We gather from Fig.~8(c) that the minority Mn 3$d$ states are completely unoccupied, while the majority states are completely occupied except for one state, resulting in a net moment of 9~$\mu_{\rm B}$ and consistent with the mixed valent states as shown schematically in Fig.~8(b).
A plot of the charge density in a small energy window corresponding to a single Mn-$d$ state in the majority spin channel (Fig.~8(c)(inset)) reveal that both the Mn ions, as well as the nearby O ions contribute to this state, indicating that the electron is hopping between both the Mn ions either directly or via O ions. Thus hopping induced interaction between the dopant Mn atoms stabilizes the tendency of FM for this system, similar to our previous observation for Fe doped ZnO cluster with the same defect state \cite{ganguli}. 

In this context, an experimental report has also pointed out that deficiency of electrons can induce ferromagnetism in Mn doped ZnO quantum dots.\cite{kittilstved} A calculation by Feng {\em et. al.} in the framework of TDDFT also claims that double exchange mechanism induced by optical excitation can stabilize ferromagnetism in Mn doped ZnO quantum dots.\cite{fengJPCL} In their description, optical excitation generates holes in the valence band of the Mn ion, thus making it behave like Mn$^{3+}$ ion. Holes thus generated participate in the double exchange process and thereby stabilize ferromagnetism.
The above discussion points to the fact that the coexistence of Mn$^{2+}$ and Mn$^{3+}$ ions in the system may be important for ferromagnetism. In the present work, we have demonstrated the role of defects to satisfy the requirement for mixed valency of Mn in the system.

Although oxygen vacancies are not very stable in the system, it has interesting effects on magnetic interactions. This is particularly visible (Fig.~7) in case of Co doping where both +1 and $-$1 charges induce ferromagnetic coupling between Co atoms. Once again, these results are similar to a recent theoretical calculation on the feromagnetism induced by O vacancies in a Co doped ZnO system \cite{sanvito08}.

\subsection{\label{sec:vacancyOnly}Defects without TM doping}
So far we have studied the role of defects in influencing the magnetic interactions between the dopant TM ions. However, recent experimental reports on ZnO nanoparticles \cite{sundaresan} and ZnO thin films \cite{ghoshal} that ZnO may be ferromagnetic even in the absence of TM dopants motivated us to investigate if defects ({\em viz.} Zn and O vacancies in different charge states) can promote ferromagnetism in ZnO cluster. For this purpose, we have calculated the formation energies corresponding to these defects and the results are displayed in Fig.~9.
\begin{figure}
\includegraphics[scale=0.45]{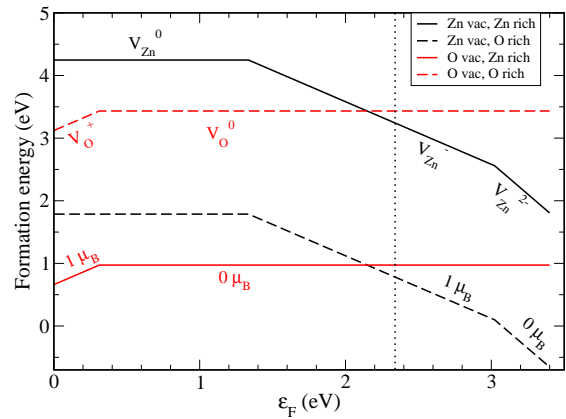}
\caption{\label{fig:vacancyOnly}(Color online) Calculated formation energies for Zn and O vacancies in the pristine ZnO cluster at Zn rich and O rich limit. Only the stable part of a particular charged state is shown. The Fermi energy $\epsilon_F$ has been varied upto the experimental band gap of bulk ZnO and the dotted vertical line indicates the calculated HOMO-LUMO gap of the pristine ZnO cluster.}
\end{figure}
We gather from this figure that O vacancies in the Zn-rich limit are stable in the charged state 0 and +1, while the Zn vacancies in the O rich limit are stable in the charged state 0, -1 and -2. A comparison of the formation energies for oxygen and Zn vacancies reveal that the most stable defects are O vacancy in the Zn rich limit in the charge state +1 (at p-type region) and also neutral O vacancies, while Zn vacancies in the O rich limit are found to be stable in the charge state -1 and -2 (at n-type region). It is interesting to note that for bulk ZnO, O vacancies in the charged state +1 exhibit negative U behavior and are never stable \cite{janotti}.  The reason for the negative U behavior is explained in Ref.~\onlinecite{janotti} and we discuss below in some details. When one oxygen atom is removed from a perfect bulk ZnO crystal, four Zn dangling bonds are created each contributing 1/2 electron to a neutral vacancy. This interaction results in a completely symmetric $a_1$ state lying in the band gap and three almost degenerate higher energy states in the conduction band. For a neutral vacancy, the $a_1$ state is occupied by two electrons and the energy is lowered when four Zn atoms surrounding the vacancy approach each other, resulting in 12\% inward relaxation \cite{janotti}. But, in case of +1 charge state of oxygen vacancy, this $a_1$ state is filled by one electron (half-filled), and the electronic energy gain is too small to overcome the strain energy, which leads to instability in the system.  On the contrary, for the cluster considered in our work, the co-ordination of O is different from that of bulk. In the cluster one oxygen atom is coordinated with three Zn atoms leading to three dangling bonds upon creation of a vacancy. For a neutral vacancy, one defect state is occupied by two electrons, which leads to large inward relaxation (6.6\%, 6.6\%, 11.8\%) similar to the bulk system \cite{janotti}. For the +1 charged state, the defect state is half filled. In a non-spin polarized calculation the Fermi level is found to shift to a peak in a narrow structured density of states (due to the reduced co-ordination in the cluster) associated with the defect.
Such a high DOS at the Fermi level is conducive for the Stoner mechanism and favors spin-polarization to lower the overall bonding energy to make the system stable resulting in a magnetic moment of 1~$\mu_B$.  So the important difference between bulk ZnO and the cluster considered in this work lies in the fact that the defect state has a reasonable width ($\sim 0.9$~eV) for bulk ZnO, which restricts the DOS at the Fermi level to a low value and does not allow gain in energy upon spin polarization. On the contrary, the defect states of the cluster with high value of DOS at the Fermi level offers gain in energy upon spin polarization \cite{coeyJPD}. The lowering of energy for oxygen vacancy in the charged state +1 is calculated to be $\sim 130$~meV upon spin-polarization concomitant with large relaxation(10.7\% outward, 10.7\% outward, 6.9\% inward) thereby adding to its stability. Similarly for the Zn vacancy the system becomes magnetic in the charge state $-1$, and therefore becomes more stable. In contrast to the bulk, Zn vacancy in the cluster in the  charged state $-1$ spans a larger energy range and the charged state $-2$ is only stable in the narrow n-type region, as shown in Fig.~9.  So the  magnetism is predominantly promoted by an unpaired electron in the charge state +1 and $-$1 for oxygen and Zn vacancies respectively and this mechanism may be responsible for magnetism in oxide nano-particles seen even in the absence of doping with magnetic elements \cite{sundaresan}.
\section{\label{sec:conclusion}Conclusion}
In conclusion, we have studied the energetics and magnetic interactions in 3d transition metal doped ZnO cluster from {\em ab initio} calculations in the framework of GGA+U. Our calculations reveal that all the 3$d$ transition metal atoms couple antiferromagnetically in a pristine ZnO cluster with the possible exception of Ti. The presence of Zn and O vacancies have crucial effect on magnetic interaction. Some of the stable defect states ({\em i.e.} vacancy with some particular charge state) are found to stabilize ferromagnetism in the cluster when TM atoms are close to each other. Such defects also stabilize unusual ionic state of the dopant TM atom at the surface of the cluster in agreement with recent experimental results. A kinetic mechanism induced by spin conserved hopping is shown to mediate ferromagnetism in the system, where the unusual ionic state of dopant TM plays an important role. We also argue that defects in ZnO clusters even in the absence of TM doping may render the cluster magnetic.

\begin{acknowledgments}
I.D. thanks DST India (No. INT/EC/MONAMI/(28)/233513/2008) for financial support. I.D. and B.S. also acknowledge Asia-Sweden Research Links Programme funded by VR/SIDA. B.S. is also grateful to Carl Tryggers Foundation for financial support.
\end{acknowledgments}


%
\end{document}